\documentclass[conference]{IEEEtran}
\usepackage{graphicx,epsfig,citesort,amsmath,amsbsy,amssymb,verbatim,url,algorithm,algpseudocode,soul}
\usepackage[usenames]{color}
\usepackage{setspace,amsfonts,epsf,url}
\usepackage{subfigure}
\usepackage{bbm}
\usepackage[normalem]{ulem}

\let\labelindent\relax
\usepackage{enumitem}
\setlist[itemize]{leftmargin=*}
\setlist[enumerate]{leftmargin=*}

\renewcommand{\baselinestretch}{0.965}
\newcommand{\x}{{\bf x}}
\newcommand{\A}{{\bf A}}
\newcommand{\y}{{\bf y}}
\newcommand{\z}{{\bf z}}
\newcommand{\f}{{\bf f}}
\newcommand{\w}{{\bf w}}
\newcommand{\n}{{\bf n}}
\newcommand{\e}{{\bf e}}

\def \l {\left}
\def \r {\right}

\newcommand{\norm}[1]{\lVert#1\rVert}

\algloopdefx{NFor}[1]{\textbf{for} #1 \textbf{do}}
\newtheorem{theorem}{Theorem}

\newcommand\asovereq{\mathrel{\stackrel{\makebox[0pt]{\mbox{\normalfont\small a.s.}}}{=}}}

\begin{document}
\title{An Overview of Multi-Processor Approximate Message Passing}

\author{\IEEEauthorblockN{Junan Zhu,\IEEEauthorrefmark{1}
Ryan Pilgrim,\IEEEauthorrefmark{2}
and
Dror Baron\IEEEauthorrefmark{2}}
\IEEEauthorblockA{\IEEEauthorrefmark{1}JPMorgan Chase \& Co., New York, NY 10001, Email: jzhu9@ncsu.edu}
\IEEEauthorblockA{\IEEEauthorrefmark{2}Department of Electrical and Computer Engineering, NC State University, Raleigh, NC 27695\\ 
Email: \{rzpilgri,barondror\}@ncsu.edu}}

\maketitle

\begin{abstract}
Approximate message passing (AMP) is an algorithmic framework for
solving linear inverse problems from noisy measurements, with exciting
applications such as reconstructing images, audio, hyper spectral images,
and various other signals, including those acquired in compressive signal acquisiton systems.
The growing prevalence of
big data systems has increased interest in large-scale problems,
which may involve huge measurement matrices that are unsuitable for conventional
computing systems. To address the challenge of large-scale processing, 
multi-processor (MP) versions of AMP have been developed.
We provide an overview of two such MP-AMP variants. In row-MP-AMP, each
computing node stores a subset of the rows of the matrix and processes corresponding
measurements. 
In column-MP-AMP, each node stores a subset of columns, and is solely
responsible for reconstructing a portion of the signal. 
We will discuss pros and cons of both approaches, 
summarize recent research results for each, and explain when
each one may be a viable approach. Aspects that are highlighted include some recent results on state evolution for  both MP-AMP algorithms, and the use of data compression to reduce communication in the MP network.
\end{abstract}


\section{Introduction}
Many scientific and engineering problems can be modeled as
solving a regularized linear inverse problem of the form
\begin{equation}
{\bf y}={\bf Ax}+{\bf w}, 
\label{eq:SP_sys} 
\end{equation}
where the goal is to estimate the unknown ${\bf x}\in\mathbb{R}^N$ given the matrix ${\bf A}\in\mathbb{R}^{M \times N}$ and statistical information about the signal ${\bf x}$ and the noise ${\bf w}\in\mathbb{R}^M$. 
These problems have received significant attention
in the compressed sensing literature~\cite{DonohoCS,CandesRUP}
with applications to image reconstruction~\cite{TMRBA2015}, communication systems~\cite{AMPMIMO2015ISIT}, and machine learning problems~\cite{Hastie2001}.

In recent years, many applications have seen explosive growth in 
the sizes of data sets. Some linear inverse problems,
for example in hyper spectral image reconstruction~\cite{Arguello2011,Arguello2013higher,TMRBA2015},
are so large that the $M\times N$ matrix elements
cannot be stored on conventional computing systems.
To solve these large-scale problems, it is possible to
partition the matrix ${\bf A}$ among multiple computing nodes
in multi-processor (MP) systems.

The matrix ${\bf A}$ can be partitioned in a column-wise or row-wise 
fashion, and the corresponding sub-matrices are stored at different processors. 
The partitioning style depends on data availability, computational considerations, and privacy concerns. 
Both types of partitioning result in reduced storage requirements per node
and faster  computation~\cite{Mota2012,Patterson2014,Han2014,Ravazzi2015,Han2015SPARS,HanZhuNiuBaron2016ICASSP,ZhuBeiramiBaron2016ISIT,ZhuBaronMPAMP2016ArXiv,MaLuBaronICASSP2017}.

{\bf Row-wise partitioning:}\
When the matrix is partitioned into rows, there 
are $P$ {\em distributed nodes} (processor nodes) and a {\em fusion center}. Each distributed node stores $\frac{M}{P}$ rows of the matrix $\A$, and acquires the corresponding linear measurements of the underlying signal $\x$. Without loss of generality, we model the measurement system in distributed node $p\in \{1,...,P\}$ as
 \begin{equation}\label{eq:one-node-meas}
    y_i={\bf a}_i \x+w_i,\ i\in \left\{\frac{M(p-1)}{P}+1,...,\frac{Mp}{P}\right\},
 \end{equation}
 where ${\bf a}_i$ is the $i$-th row of $\A$, and $y_i$ and $w_i$ are the $i$-th entries of $\y$ and $\w$, respectively.
Once every $y_i$ is collected, we run distributed algorithms among the fusion center and $P$ distributed nodes to reconstruct the signal $\x$. Prior studies on solving row-wise partitioned linear inverse problems include extending existing algorithms such as least absolute shrinkage and selection operator 
(LASSO)~\cite{Hastie2001} and iterative hard thresholding (IHT)  
to a distributed setting~\cite{Mota2012,Han2015SPARS}.

{\bf Column-wise partitioning:}\
Columns of the matrix ${\bf A}$ may correspond to features in feature selection problems~\cite{Hastie2001}.
In some applications, for example in healthcare when rows of the matrix correspond to patients,
privacy concerns or other constraints prevent us from storing
entire rows (corresponding to all the data about a patient) in individual processors, 
and column-wise partitioning becomes preferable.
The (non-overlapping) column-wise partitioned linear inverse problem
can be modeled as follows,
\begin{equation}
{\bf y}=\sum_{p=1}^P{\bf A}^p{\bf x}^p+{\bf w}, 
\label{eq:MP_sys}
\end{equation}
where ${\bf A}^p\in\mathbb{R}^{M\times N_p}$ is the sub-matrix that is stored in processor $p$, and $\sum_{p=1}^P N_p=N$.

Many studies 
on solving the column-wise partitioned linear inverse problem~(\ref{eq:MP_sys})
have been in the context of distributed feature selection. For example,
Zhou {\em et al.}~\cite{Zhou2014} modeled feature selection as a parallel group testing problem.
Wang {\em et al.}~\cite{WangDunsonLeng2016} proposed to de-correlate the data matrix before partitioning, 
so that each processor can work independently using the de-correlated matrix without communication with other processors.
Peng {\em et al.}~\cite{PengYanYin2013} studied problem~(\ref{eq:MP_sys}) in the context of optimization, where they proposed a greedy coordinate-block descent algorithm and a parallel implementation of 
the fast iterative shrinkage-thresholding algorithm
(FISTA)~\cite{Beck2009FISTA}.

This paper relies on approximate message passing
(AMP)~\cite{DMM2009,Montanari2012,Bayati2011,Krzakala2012probabilistic},
an iterative framework that solves linear inverse problems.
We overview the recent progress in 
understanding the distributed AMP algorithm applied to either row-wise or column-wise partitioned linear inverse problems.

The rest of the paper is organized as follows.
After reviewing the AMP literature in Section~\ref{sec:background}, Section~\ref{sec:MP-CS_for_MP-AMP} discusses the row-partitioned version,
and the column-partitioned version appears in Section~\ref{sec:column_MP-AMP}.
We conclude the paper in Section~\ref{sec:discussion}.


\section{Approximate Message Passing}\label{sec:background}

To solve large-scale MP linear inverse problems partitioned 
either row-wise or column-wise, we use 
approximate message passing (AMP)~\cite{DMM2009,Montanari2012,Bayati2011,Krzakala2012probabilistic},
an iterative framework that solves linear inverse problems by successively decoupling~\cite{Tanaka2002,GuoVerdu2005,GuoWang2008} matrix channel problems into scalar channel denoising
problems with additive white Gaussian noise (AWGN). AMP has received considerable attention because of its fast convergence, computational efficiency, and state evolution (SE) formalism~\cite{DMM2009,Bayati2011,Rush_ISIT2016}, which offers a precise
characterization of the AWGN denoising problem in each iteration.
In the Bayesian setting, AMP often achieves the minimum mean squared error (MMSE)~\cite{ZhuBaronCISS2013,Krzakala2012probabilistic} in the limit of large linear systems. Various extensions to AMP have been considered since AMP was initially introduced. Below, we summarize recent developments in AMP theory and application.

\textbf{Generalizations of AMP:}
Recently, a number of authors have studied the incorporation of various non-separable denoisers within AMP~\cite{Tan_CompressiveImage2014,TMRBA2015,Metzler2016_IEEE,MaZhuBaron2016TSP,JavanmardMontanari2012,Rush_ISIT2015}, generalization of the measurement matrix prior~\cite{Swamp2014,Vila2015,RanganADMMGAMP2015_ISIT,SAMP_nonlinear2015,VAMP2016}, and  relaxation of assumptions on the probabilistic observation model~\cite{RanganGAMP2011ISIT,JavanmardMontanari2012,SAMP_nonlinear2015}. AMP-based methods have also been applied to solve the bilinear inference problem~\cite{BiGAMP1,BiGAMP2,Kabashima2016}, with 
matrix factorization applications. 

{\bf Applications:} The AMP framework and its many extensions have found applications in capacity-achieving sparse superposition codes~\cite{Rush_ISIT2015}, compressive imaging~\cite{Som2012,Tan_CompressiveImage2014,Metzler2016_IEEE}, 
hyperspectral image reconstruction~\cite{TMRBA2015} and hyperspectral unmixing~\cite{VilaSchniterMeola2015_IEEE}, universal compressed sensing reconstruction~\cite{MaZhuBaron2016TSP}, MIMO detection~\cite{AMPMIMO2015ISIT},
and matrix factorization applications~\cite{BiGAMP1,BiGAMP2,Kabashima2016}.

{\bf Multi-processor AMP:}
Recently, Zhu {\em et al.}~\cite{ZhuBeiramiBaron2016ISIT,ZhuBaronMPAMP2016ArXiv} studied the application of lossy compression in row-wise partitioned MP-AMP, such that the cost of running the reconstruction algorithm is minimized. Ma {\em et al.}~\cite{MaLuBaronICASSP2017} proposed a distributed version of AMP to solve column-wise partitioned linear inverse problems, with a rigorous study of
state evolution. 

{\bf Centralized AMP:}\
Our model for the linear system~\eqref{eq:SP_sys}
includes an independent and identically distributed (i.i.d.) 
Gaussian measurement matrix $\A$, i.e., $A_{i,j}\sim\mathcal{N}(0,\frac{1}{M})$.\footnote{
When the matrix $\A$ is not i.i.d. Gaussian, the use of damping or other variants of AMP algorithms such as Swept AMP~\cite{Swamp2014} and VAMP~\cite{VAMP2016} is necessary in order for the algorithm to converge. This paper only considers
an i.i.d. Gaussian matrix $\A$ in order
to present some theoretical results; the theoretic understanding of using AMP in general matrices is less mature.}
The signal entries follow an i.i.d. distribution.
The noise entries obey $w_i\sim\mathcal{N}(0,\sigma_W^2)$, where $\sigma_W^2$ is the noise variance.

Starting from ${\bf x}_0={\bf 0}$ and ${\bf z}_0={\bf 0}$,  the AMP framework~\cite{DMM2009} proceeds iteratively according to
\begin{align}
{\bf x}_{t+1}&=\eta_t({\bf A}^{T}{\bf z}_t+{\bf x}_t)\label{eq:AMPiter1},\\
{\bf z}_t&={\bf y}-{\bf Ax}_t+\frac{1}{\kappa}{\bf z}_{t-1}
\langle d\eta_{t-1}({\bf A}^{T}{\bf z}_{t-1}+{\bf x}_{t-1})\rangle\label{eq:AMPiter2},
\end{align}
where $\eta_t(\cdot)$ is a denoising function, $d\eta_{t}(\cdot)=\frac{d \eta_t({\cdot})}{d\{\cdot\}}$ is shorthand for the derivative of $\eta_t(\cdot)$, and~$\langle{\bf u}\rangle=\frac{1}{N}\sum_{i=1}^N u_i$
for some vector~${\bf u}\in\mathbb{R}^N$. The subscript $t$ represents the iteration index, $T$ denotes transpose, and $\kappa=\frac{M}{N}$ is the measurement rate.
Owing to the decoupling effect~\cite{Tanaka2002,GuoVerdu2005,GuoWang2008}, in each AMP iteration~\cite{Bayati2011,Montanari2012},
the vector~$\f_t={\bf A}^{T}{\bf z}_t+{\bf x}_t$
in (\ref{eq:AMPiter1}) is statistically equivalent to
the input signal ${\bf x}$ corrupted by AWGN ${\bf e}_t$ generated by a source $E\sim \mathcal{N}(0,\sigma_t^2)$,
\begin{equation}\label{eq:equivalent_scalar_channel}
\f_t=\x+{\bf e}_t.
\end{equation}
In large systems ($N\rightarrow\infty, \frac{M}{N}\rightarrow \kappa$), a useful property of AMP~\cite{Bayati2011,Montanari2012} is that
the noise variance $\sigma_t^2$ of the equivalent scalar channel~\eqref{eq:equivalent_scalar_channel} evolves following SE:
\begin{equation}\label{eq:SE_centralized}
\sigma_{t+1}^2=\sigma^2_W+\frac{1}{\kappa}\text{MSE}(\eta_t,\sigma_t^2),
\end{equation}
where the mean squared error (MSE) is
$\text{MSE}(\eta_t,\sigma_t^2)=\mathbb{E}_{X,E}\left[\left( \eta_t\left( X+E \right)-X \right)^2\right]$, $\mathbb{E}_{X,W}(\cdot)$ is expectation with respect to $X$ and $E$, and $X\sim f_X$ is the source that generates $\x$.
Formal statements for SE appear
in prior work~\cite{Bayati2011,Montanari2012,Rush_ISIT2016}.

The SE in~\eqref{eq:SE_centralized} can also be expressed in the following recursion,
\begin{align}
\tau_t^2 &=\sigma_W^2 +  \sigma_t^2,\nonumber\\
\sigma_{t+1}^2 &=\kappa^{-1}\mathbb{E}\left[\left(\eta_t(X+\tau_{t}Z)-X\right)^2\right],\label{eq:SP_SE}
\end{align} 
where $Z$ is a standard normal random variable (RV) 
that is independent of $X$, and $\sigma_0^2=\kappa^{-1}\mathbb{E}[X^2]$.

This paper considers the Bayesian setting, which assumes knowledge of the true prior for the signal $\x$. Therefore, the MMSE-achieving denoiser is conditional expectation, $\eta_t(\cdot)=\mathbb{E}[\x|\f_t]$, which is easily obtained.
Other denoisers such as soft thresholding~\cite{DMM2009,Montanari2012,Bayati2011} yield MSE's  that are greater than that of the Bayesian denoiser.
When the true prior for $\x$ is unavailable, parameter estimation techniques
can be used~\cite{EMGMTSP,Kamilov2014,MaZhuBaron2016TSP}.

\section{Row-wise MP-AMP}\label{sec:MP-CS_for_MP-AMP}

\subsection{Lossless R-MP-AMP}

Han {\em et al.}~\cite{Han2015ICASSP} proposed AMP for row-wise partitioned MP linear inverse problems (R-MP-AMP) for a network with $P$ processor nodes and a fusion center. 
Each processor node stores rows of the matrix $\A$ as in~\eqref{eq:one-node-meas}, 
carries out the decoupling step of AMP, and generates part of the pseudo data $\f_t^p$. The fusion center merges the pseudo data sent by all processor nodes, 
$\f_t=\sum_{p=1}^P \f_t^p$, denoises $\f_t$, and sends back the denoised $\f_t$ to each processor node. The detailed steps are summarized in Algorithm~\ref{algo:R-MP-AMP-lossless}. Mathematically, Algorithm~\ref{algo:R-MP-AMP-lossless} is equivalent to the centralized AMP in~\eqref{eq:AMPiter1}-\eqref{eq:AMPiter2}. Therefore, the SE in~\eqref{eq:SE_centralized} tracks the evolution of Algorithm~\ref{algo:R-MP-AMP-lossless}. Note that 
${\bf a}^p$ denotes the row partition of the matrix $\A$ at processor $p$.

\begin{algorithm}
\caption{R-MP-AMP (lossless)}\label{algo:R-MP-AMP-lossless}
\textbf{Inputs to Processor $p$:} ${\bf y}$, ${\bf a}^p$, $\widehat{t}$\\
\textbf{Initialization:}  ${\bf x}_{0}={\bf 0},\ {\bf z}_{0}^p={\bf 0}, \forall p$
\begin{algorithmic}
\NFor{$t=1:\widehat{t}$}\\
$\quad$At Processor $p$:\\
$\quad {\bf z}_t^p={\bf y}^p-{\bf a}^p\x_t+\frac{1}{\kappa}{\bf z}_{t-1}^p
g_{t-1},\quad {\bf f}_t^p=\frac{1}{P}\x_t+({\bf a}^p)^{T}{\bf z}_t^p$\\
$\quad$At fusion center:\\
$\quad {\bf f}_{t}=\sum_{p=1}^P {\bf f}_{t}^p,\quad g_{t}=\langle d\eta_{t}({\bf f}_{t})\rangle,\quad \x_{t+1}=\eta_{t}( {\bf f}_{t})$
\end{algorithmic}
\textbf{Output from fusion center:} ${\bf x}_{\widehat{t}}$
\end{algorithm}

\subsection{Lossy R-MP-AMP}

In lossless R-MP-AMP (Algorithm~\ref{algo:R-MP-AMP-lossless}), 
the processor nodes and fusion center send real-valued vectors 
of length $N$ to each other, i.e., $\f_t^p$ and $\x_{t+1}$, at
floating point precision. However, in some applications it is costly to send uncompressed real numbers
at full precision.
To reduce the communication load of inter-node messages, 
we use lossy compression~\cite{Cover06,Berger71}. 

Applying lossy compression to the messages sent from each processor node to the fusion center, we obtain the
lossy R-MP-AMP~\cite{HanZhuNiuBaron2016ICASSP,ZhuBaronMPAMP2016ArXiv} steps as described in Algorithm~\ref{algo:R-MP-AMP-lossy}, where $Q(\cdot)$ denotes quantization.

\begin{algorithm}
\caption{R-MP-AMP (lossy)}\label{algo:R-MP-AMP-lossy}
\textbf{Inputs to Processor $p$:} ${\bf y}$, ${\bf a}^p$, $\widehat{t}$\\
\textbf{Initialization:}  ${\bf x}_0={\bf 0},\ {\bf z}_{0}^p={\bf 0}, \forall p$
\begin{algorithmic}
\NFor{$t=1:\widehat{t}$}\\
$\quad$At Processor $p$:\\
$\quad {\bf z}_t^p={\bf y}^p-{\bf a}^p\x_t+\frac{1}{\kappa}{\bf z}_{t-1}^p
g_{t-1},\quad {\bf f}_t^p=\frac{1}{P}\x_t+({\bf a}^p)^{T}{\bf z}_t^p$\\
$\quad$At fusion center:\\
$\quad {\bf f}_{Q,t}=\sum_{p=1}^P Q({\bf f}_{t}^p),\quad g_{t}=\langle d\eta_{t}({\bf f}_{Q,t})\rangle$,\\ 
$\quad \x_{t+1}=\eta_{t}( {\bf f}_{Q,t})$
\end{algorithmic}
\textbf{Output from fusion center:} ${\bf x}_{\widehat{t}}$
\end{algorithm}

The reader might notice that the fusion center also needs to transmit the denoised signal vector $\x_t$ and a scalar $g_{t-1}$ to the distributed nodes. The transmission of the scalar $g_{t-1}$ is negligible relative to the transmission of $\x_t$, and the fusion center may broadcast $\x_t$ so that naive compression of $\x_t$, such as compression with a fixed quantizer, is sufficient. Hence, we will not discuss possible lossy compression of the messages transmitted by the fusion center. 

Assume that we quantize $\f_t^p, \forall p$, and use $C$ bits on average
to encode the quantized vector $Q(\f_t^p)\in\mathcal{X}^N \subset \mathbb{R}^N$, 
where $\mathcal{X}$ is a set of representation levels. The per-symbol {\em coding rate} is $R=\frac{C}{N}$. We incur an {\em expected distortion}
\begin{equation*}
D_t^p=\mathbb{E}\left[\frac{1}{N}\sum_{i=1}^N(Q(f_{t,i}^p)-f_{t,i}^p)^2\right]
\end{equation*}
at iteration $t$ in each processor node,\footnote{Because we assume that
$\A$ and $\z$ are both i.i.d., the expected distortions are the same over
all $P$ nodes, and can be denoted by $D_t$ for simplicity.
Note also that $D_t=\mathbb{E}[(Q(f_{t,i}^p)-f_{t,i}^p)^2]$
due to $\x$ being i.i.d.}
where $Q(f_{t,i}^p)$ and $f_{t,i}^p$ are the $i$-th entries of the vectors $Q(\f_t^p)$ and $\f_t^p$, respectively,
and expectation is over $\f_t^p$.
When the size of the problem grows, i.e., $N\rightarrow\infty$, the rate-distortion (RD) function, denoted by $R(D)$, offers the 
information theoretic limit on the coding rate $R$ for communicating a long sequence up to distortion $D$~\cite{Cover06,Berger71,GershoGray1993}.
A pivotal conclusion from RD theory is that coding rates can be greatly reduced even if $D$ is small.
The function $R(D)$ can be computed in various ways~\cite{Arimoto72,Blahut72,Rose94} 
and can be achieved by an RD-optimal quantization scheme in the limit of large $N$.
Other quantization schemes will require larger coding rates to achieve the same expected distortion $D$.

Assume that
appropriate vector quantization (VQ) schemes~\cite{LBG1980,Gray1984,GershoGray1993} that achieve $R(D)$ are applied within each MP-AMP iteration. The signal {\em at the fusion center} before denoising can then be modeled as
\begin{align}
\f_{Q,t}=\sum_{p=1}^P Q(\f_t^p)=\x+{\bf e}_t+{\bf n}_t,\label{eq:indpt_noises}
\end{align}
where ${\bf e}_t$ is the equivalent scalar channel noise~\eqref{eq:equivalent_scalar_channel} and ${\bf n}_t$ is the overall quantization error. For large block sizes, we expect the VQ quantization error ${\bf n}_t$ to resemble additive white Gaussian noise with variance $PD_t$ that is independent of $\x+{\bf e}_t$ at high rates, or at all rates using dithering~\cite{ZamirFeder96}.

{\bf State evolution for lossy R-MP-AMP:}
Han {\em et al.} suggest that SE for lossy R-MP-AMP~\cite{HanZhuNiuBaron2016ICASSP} follows 
\begin{equation}
\sigma_{t+1}^2=\sigma^2_W+\frac{1}{\kappa}\text{MSE}(\eta_t,\sigma_t^2+PD_t),\label{eq:SE_Q}
\end{equation}
where $\sigma_t^2$ can be estimated by
$\widehat{\sigma}_t^2 = \frac{1}{M}\|{\bf z}_t\|_2^2$ with $\|\cdot\|_p$ denoting the $\ell_p$ norm~\cite{Bayati2011,Montanari2012}, and $\sigma_{t+1}^2$ is the variance of ${\bf e}_{t+1}$.
The rigorous justification of~\eqref{eq:SE_Q} by extending the framework put forth by Bayati and Montanari~\cite{Bayati2011} and Rush and Venkataramanan~\cite{Rush_ISIT2016_arxiv} is left for future work. 
Instead, we argue that lossy SE~\eqref{eq:SE_Q} asymptotically tracks the evolution of $\sigma_t^2$ in lossy MP-AMP in the limit of low normalized distortion $\frac{PD_t}{\sigma_t^2}\rightarrow 0$.
Our argument is comprised of three parts: ({\em i}) $\e_t$ and $\n_t$~\eqref{eq:indpt_noises} are approximately independent in the limit of $\frac{PD_t}{\sigma_t^2}\rightarrow 0$,   ({\em ii})  $\e_t+\n_t$ is approximately independent of $\x$ in the limit of $\frac{PD_t}{\sigma_t^2}\rightarrow 0$, and ({\em iii}) lossy SE~\eqref{eq:SE_Q} holds if ({\em i}) and ({\em ii}) hold.
The first part ($\e_t$ and $\n_t$ are independent) ensures that we can track the variance of $\e_t+\n_t$ with $\sigma_t^2+PD_t$. The second part ($\e_t+\n_t$ is independent of $\x$) ensures that lossy MP-AMP
follows lossy SE~\eqref{eq:SE_Q} as it falls under the general framework discussed in Bayati and Montanari~\cite{Bayati2011} and Rush and Venkataramanan~\cite{Rush_ISIT2016_arxiv}. Hence, the third part of our argument holds. The numerical justification of these three parts appears in Zhu {\em et al.}~\cite{ZhuBaronMPAMP2016ArXiv,ZhuDissertation2017}.

{\bf Optimal coding rates:}
Denote the coding rate used to transmit $Q(\f^p_t)$ at iteration $t$ by $R_t$. The sequence of $R_t,\ t=1,...,\widehat{t}$, where $\widehat{t}$ is the total number of MP-AMP iterations, is called
the {\em coding rate sequence}, and is denoted by the vector $\mathbf{R}=[R_1,..., R_{\widehat{t}}]$. Given the coding rate sequence $\mathbf{R}$, the distortion $D_t$ can be evaluated with $R(D)$, and the scalar channel noise variance $\sigma_t^2$ can be evaluated with~\eqref{eq:SE_Q}.
Hence, the MSE for $\mathbf{R}$ can be predicted.
The coding rate sequence $\mathbf{R}$ can be optimized using dynamic programming (DP)~\cite{CLR,ZhuBaronMPAMP2016ArXiv}.
That said, our recent theoretical analysis of lossy R-MP-AMP has revealed that
the coding rate is linear in the limit of EMSE$\rightarrow 0$, where EMSE denotes excess MSE (EMSE=MSE-MMSE).
This result is summarized in the following theorem.

\begin{theorem}[Linearity of the coding rate sequence~\cite{ZhuBaronMPAMP2016ArXiv}]\label{th:optRateLinear}
Supposing that lossy SE~\eqref{eq:SE_Q} holds, we have 
\begin{equation*}
\lim_{t \to \infty} \frac{D_{t+1}^*}{D_t^*} = \theta,
\end{equation*}
where $\theta=\frac{N}{M}\text{MSE}'(\sigma_{\infty}^2)$ and $D_{t}^*$ denotes the optimal distortion at iteration $t$. Further, define the additive growth rate at iteration $t$ as $R_{t+1}-R_t$. 
The additive growth rate for the optimal coding rate sequence $\mathbf{R}^*$ 
satisfies 
\begin{equation*}
\lim_{t\rightarrow\infty} \l(R^*_{t+1} - R^*_{t }\r)= \frac{1}{2}\log_2 \l(\frac{1}{\theta}\r).
\end{equation*}
\end{theorem}

\begin{figure*}[t!]
  \centering
  \begin{minipage}[t]{0.31\textwidth}
    \centering
    \includegraphics[width=1\textwidth]{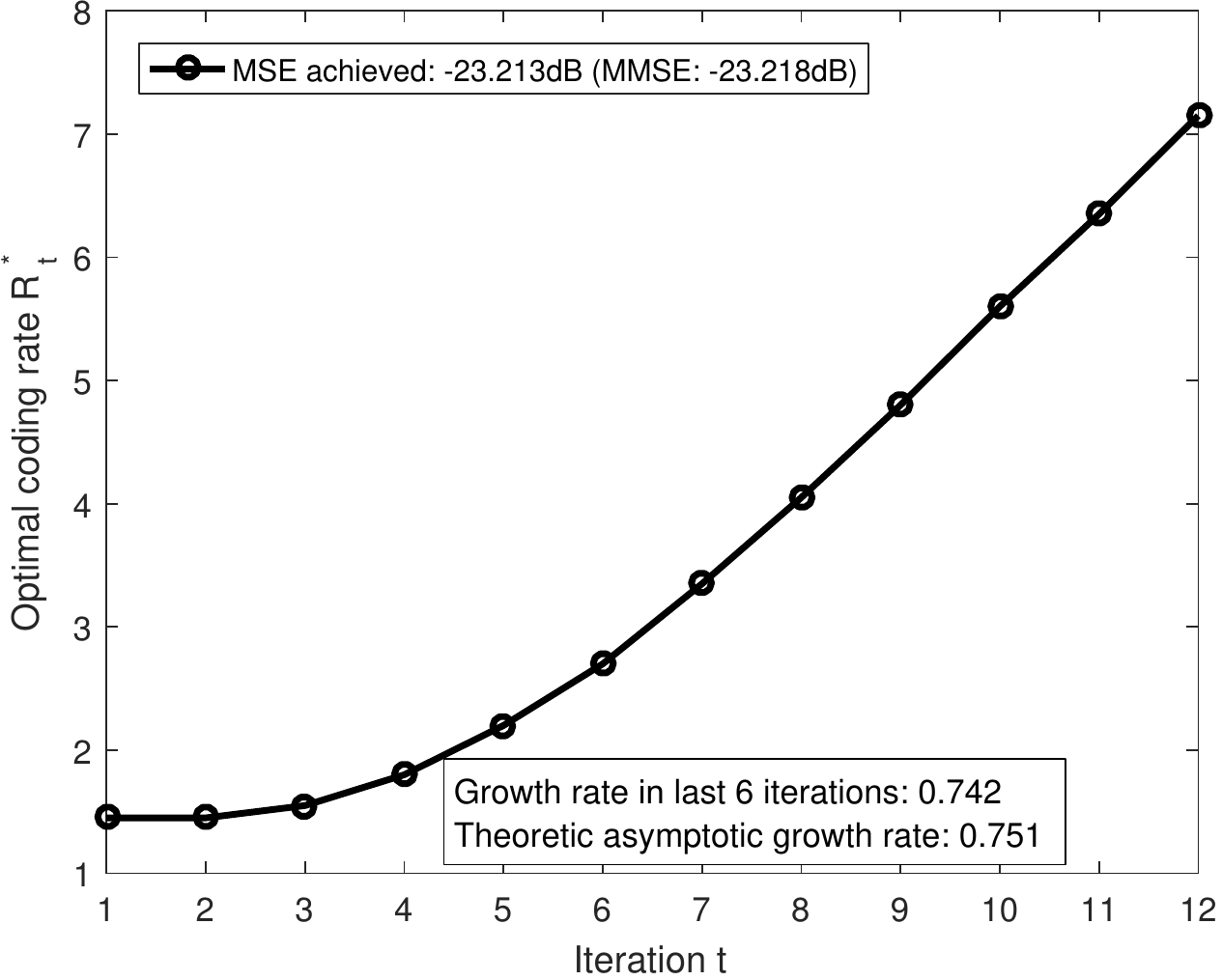}
    \caption{
Low-EMSE growth rate of optimal coding rate sequence per DP
vs. asymptotic growth rate $\frac{1}{2}\log_2\l(\frac{1}{\theta}\r)$. (BG signal~\eqref{eq:BG}, $\rho=0.2,\ \kappa=1,\ P=100, \sigma_W^2=0.01,\ b=0.782$.)
}\label{fig:asympSlope}
  \end{minipage}\hfill
  \begin{minipage}[t]{0.32\textwidth}
  	\centering
    \includegraphics[width=1\textwidth]{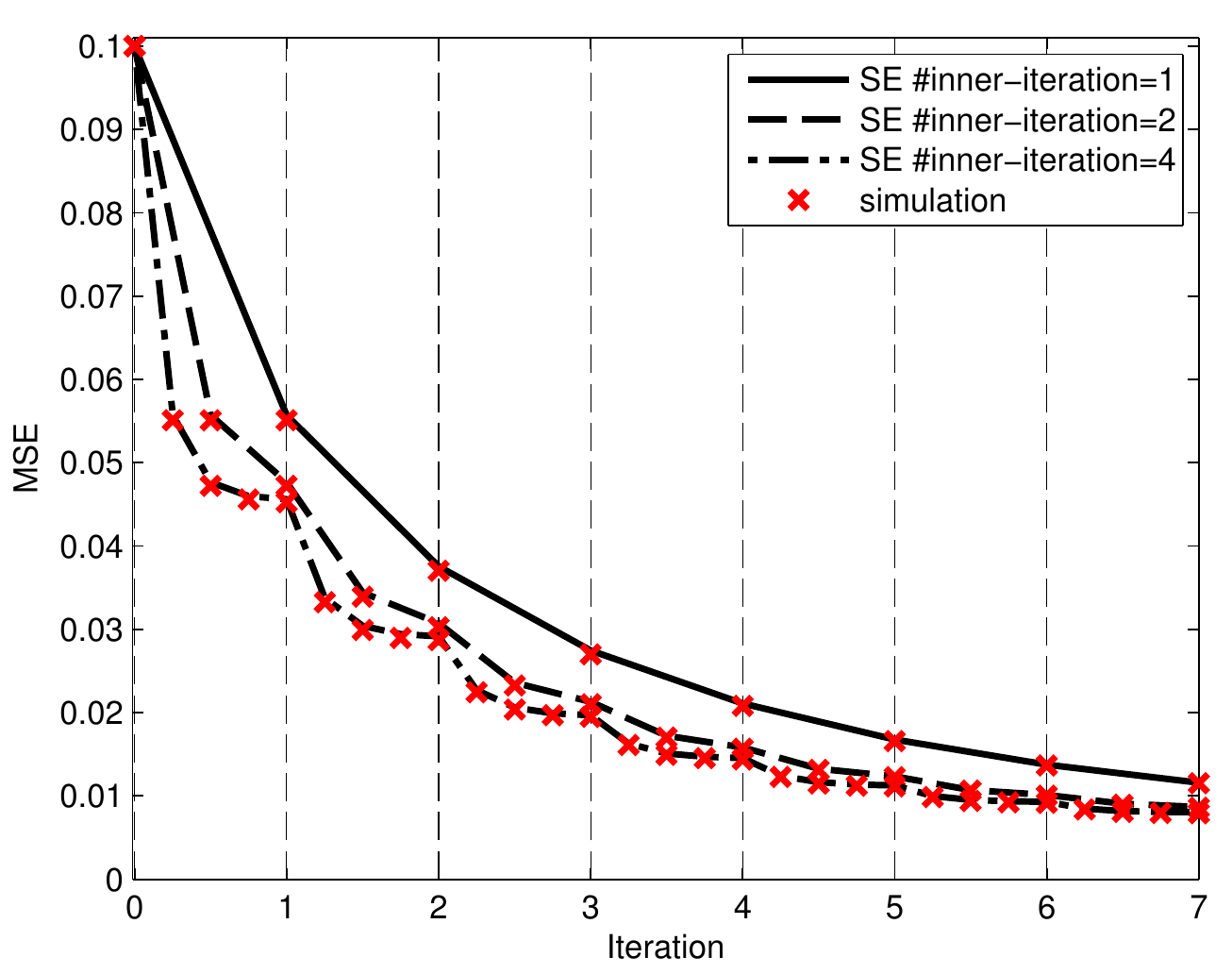}
    \caption{Verification of SE for C-MP-AMP with various communication schedules. (P=3, N=30000, M=9000, SNR=15dB.)}\label{fig:SE}
  \end{minipage}\hfill
  \begin{minipage}[t]{0.32\textwidth}
    \centering
    \includegraphics[width=1\textwidth]{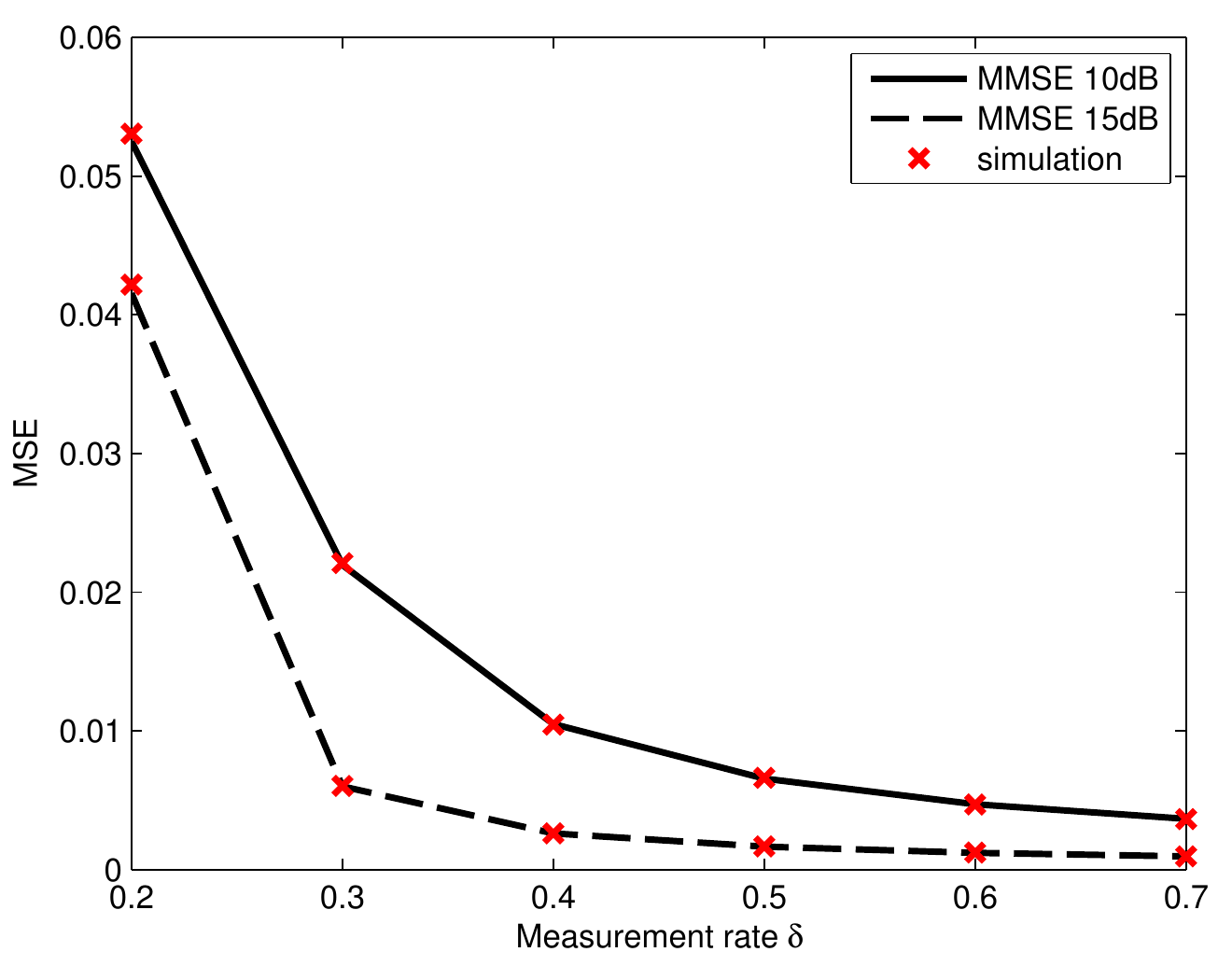}
    \caption{Verification that C-MP-AMP achieves the MMSE at various measurement rates $\kappa=M/N$ and SNR levels. (P=3, N=30000.)}\label{fig:MMSE}
  \end{minipage}
  \vspace*{-4mm}
\end{figure*}

{\bf Comparison of DP results to Theorem~\ref{th:optRateLinear}:}
We run DP (discussed in Zhu {\em et al.}~\cite{ZhuBaronMPAMP2016ArXiv}) to find an optimal coding rate sequence $\mathbf{R}^*$ 
to reconstruct a {\em Bernoulli-Gaussian} (BG) signal, whose entries follow
\begin{equation}
x_j\sim 
\rho \mathcal{N}(0,1)+(1-\rho)\delta(x_j),\label{eq:BG}
\end{equation}
where $\delta(\cdot)$ is the Dirac delta function\, and $\rho$ is called the {\em sparsity rate} of the signal. The detailed setting is: sparsity rate $\rho=0.2$, $P=100$ nodes, measurement rate $\kappa=1$, noise variance $\sigma_W^2=0.01$, and 
normalized cost ratio of computation to communication $b=0.782$ (a formal definition of $b$ appears 
in~\cite{ZhuBaronMPAMP2016ArXiv}). 
The goal is to achieve a desired EMSE of 0.005 dB, i.e.,
$10\log_{10}\l(1 + \frac{\text{EMSE}}{\text{MMSE}}\r)=0.005$.
We use uniform ECSQ~\cite{GershoGray1993,Cover06} with optimal block entropy coding~\cite{GershoGray1993} at each processor node 
and the corresponding relation between the
rate $R_t$ and distortion $D_t$ of ECSQ in the DP scheme. We know that ECSQ achieves a coding rate
within an additive constant of the RD function $R(D)$ at high rates~\cite{GershoGray1993}. Therefore, the additive
growth rate of the optimal coding rate sequence
obtained for ECSQ will be the same as the additive growth rate if the RD relation is modeled by $R(D)$~\cite{Cover06,Berger71,GershoGray1993}.

The resulting optimal coding rate sequence is plotted in Fig.~\ref{fig:asympSlope}. The additive growth rate of the last six iterations is $\frac{1}{6}(R_{12}^*-R_{6}^*)=0.742$, while the asymptotic additive growth rate according to Theorem~\ref{th:optRateLinear} is $\frac{1}{2}\log_2\l(\frac{1}{\theta}\r)\approx 0.751$.
Note that the discrepancy of 0.009 between the additive growth rate from the simulation and the asymptotic additive growth rate is
within the numerical precision of our DP scheme.
In conclusion, our numerical result matches the theoretical prediction of Theorem~\ref{th:optRateLinear}.

\begin{algorithm}
\caption{C-MP-AMP (lossless)}\label{algo:C-MP-AMP}
\textbf{Inputs to Processor $p$:} ${\bf y}$, ${\bf A}^p$, $\{\widehat{t}_s\}_{s=0,...,\widehat{s}}$ (maximum number of inner iterations at each outer iteration)\\
\textbf{Initialization:}  ${\bf x}^p_{0,\widehat{t}_0}={\bf 0}$, ${\bf z}^p_{0,\widehat{t}_0-1}={\bf 0}$, ${\bf r}^p_{0,\widehat{t}_0}={\bf 0}$, $\forall p$
\begin{algorithmic}
\NFor{$s=1:\widehat{s}$} (loop over outer iterations)\\
$\:$At fusion center:
$\:{\bf g}_{s}=\sum_{u=1}^{P}{\bf r}^u_{s-1,\widehat{t}_{s-1}}$\\
$\:$At Processor $p$:\\
$\:{\bf x}^p_{s,0}={\bf x}^p_{s-1,\widehat{t}_{s-1}}$, ${\bf r}^p_{s,0}={\bf r}^p_{s-1,\widehat{t}_{s-1}}$
\NFor{$t=0:\widehat{t}_s-1$} (loop over inner iterations)\\
$\hspace{0.24in} {\bf z}^p_{s,t}={\bf y}-{\bf g}_s-\left({\bf r}^p_{s,t}-{\bf r}^p_{s,0}\right)$\\
$\hspace{0.24in}{\bf x}^p_{s,t+1}=\eta_{s,t}({\bf x}^p_{s,t}+({\bf A}^p)^T {\bf z}^p_{s,t})$\\
$\hspace{0.24in}{\bf r}^p_{s,t+1}={\bf A}^p {\bf x}^p_{s,t+1}-\frac{{\bf z}^p_{s,t}}{M}\sum_{i=1}^{N_p}\eta_{s,t}'([{\bf x}^p_{s,t}+({\bf A}^p)^T{\bf z}^p_{s,t}]_i)$
\end{algorithmic}
\textbf{Output from processor $p$:} ${\bf x}^p_{\widehat{s},\widehat{t}_{\widehat{s}}}$
\end{algorithm}

\section{Column-wise MP-AMP}
\label{sec:column_MP-AMP}
In our proposed column-wise multiprocessor AMP (C-MP-AMP) algorithm~\cite{MaLuBaronICASSP2017},
the fusion center collects vectors that represent the estimates 
of the portion of the measurement vector ${\bf y}$
contributed by the data from individual processors.
The sum of these vectors is computed in the fusion center and transmitted to all processors.
Each processor performs standard AMP iterations with a new equivalent measurement vector, which
is computed using the vector received from the fusion center.
The pseudocode for C-MP-AMP is presented in Algorithm~\ref{algo:C-MP-AMP}.

{\bf State evolution:}
Similar to AMP, the dynamics of the C-MP-AMP algorithm can be characterized by an SE formula. 
Let $(\sigma_{0,\widehat{t}}^p)^2 =\kappa_p^{-1}\mathbb{E}[X^2]$, where $\kappa_p=M/N_p$, $\forall p=1,...,P$. 
For outer iterations $1\leq s\leq \widehat{s}$ and inner iterations $0\leq t\leq\widehat{t}_s$, we define the sequences $\{(\sigma^p_{s,t})^2\}$ and $\{(\tau^p_{s,t})^2\}$ as
\begin{align}
(\sigma^p_{s,0})^2&=(\sigma^p_{s-1,\widehat{t}})^2,\label{eq:MP_SE1}\\
(\tau^p_{s,t})^2 &=\sigma_W^2+\sum_{u=1}^{P}(\sigma^u_{s,0})^2+\left((\sigma^p_{s,t})^2-(\sigma^p_{s,0})^2\right),\label{eq:MP_SE2}\\
(\sigma^p_{s,t+1})^2 &=\kappa_p^{-1}\mathbb{E}\left[\left(\eta_{s,t}(X+\tau^p_{s,t}Z)-X\right)^2 \right],\label{eq:MP_SE3}
\end{align}
where $Z$ is standard normal and independent of $X$. 
With these definitions, we have the following theorem for C-MP-AMP.

\begin{theorem}[\cite{MaLuBaronICASSP2017}]\label{th:SE_C_MP_AMP}
Under the assumptions listed in \cite[Section 1.1]{Rush_ISIT2016_arxiv}, for $p=1,...,P$, let $M/N_p\rightarrow\kappa_p\in (0,\infty)$ be a constant. Define $N=\sum_{p=1}^P N_p$. Then for any PL(2) function\footnote{A function $f: \mathbb{R}^{m} \to \mathbb{R}$ is \emph{pseudo-Lipschitz} of order-2, denoted PL(2), if there exists a constant $L >0$ such that for all $x, y \in \mathbb{R}^{m}$, $|\phi(x)-\phi(y)|\leq L(1+\norm{x}+\norm{y})\norm{x-y}$, where $\norm{\cdot}$ denotes the Euclidean norm.} $\phi:\mathbb{R}^2\rightarrow\mathbb{R}$, we have
\begin{equation*}
\lim_{N\rightarrow\infty}\frac{1}{N_p}\!\sum_{i=1}^{N_p}\!\phi([x^{p}_{s,t+1}]_i,x^{p}_i)\asovereq\mathbb{E}\left[\phi(\eta_{s,t}(X+\tau^p_{s,t}Z),X)\right]\!,\!\forall p,
\end{equation*}
where ${\bf x}^p_{s,t+1}$ is generated by the C-MP-AMP algorithm, $\tau^p_{s,t}$ is defined in (\ref{eq:MP_SE1}--\ref{eq:MP_SE3}), 
$x^{p}_i$ is the $i^{\text{th}}$ element in $x^p$, 
$x^p$ is the true signal in the $p^{\text{th}}$ processor,
$X\sim p_X$, and $Z$ is a standard normal RV that is independent of $X$.
\end{theorem}  

{\em Remark 1: C-MP-AMP converges to a fixed point that is no worse than that of AMP.} This statement can be demonstrated as follows.
When C-MP-AMP converges, the quantities in (\ref{eq:MP_SE1}--\ref{eq:MP_SE3}) do not keep changing, hence we can drop all the iteration indices for fixed point analysis. 
Notice that the last term on the right hand side (RHS) of (\ref{eq:MP_SE2}) vanishes, which leaves the RHS independent of $p$. 
Denote $(\tau^p_{s,t})^2$ by $\tau^2$ for all $s,t,p$, and plug (\ref{eq:MP_SE3}) into (\ref{eq:MP_SE2}), then
\begin{align*}
\tau^2&=\sigma_W^2+\sum_{p=1}^{P}\kappa_p^{-1}\mathbb{E}\left[\left(\eta(X+\tau Z)-X\right)^2\right]\\
&\overset{(a)}{=}\sigma_W^2+\kappa^{-1}\mathbb{E}\left[\left(\eta(X+\tau Z)-X\right)^2\right],
\end{align*}
which is identical to the fixed point equation obtained from
(\ref{eq:SP_SE}), where (a) holds because $\sum_{p=1}^P \kappa_p^{-1}=\sum_{p=1}^P \frac{N_p}{M} = \frac{N}{M}.$
Because AMP always converges to the worst fixed point of
(\ref{eq:SP_SE})~\cite{Krzakala2012probabilistic}, 
the average asymptotic performance of C-MP-AMP 
is at least as good as AMP.


{\em Remark 2: The asymptotic dynamics of C-MP-AMP can be identical to AMP with a specific communication schedule.} This can be achieved by letting
$\widehat{t}_s=1,\forall s$. In this case, the quantity $(\tau^p_{s,t})$ is involved only for $t=0$. 
Because the last term in (\ref{eq:MP_SE2}) is 0 when $t=0$, the computation
of $(\tau^p_{s,0})^2$ is independent of $p$. 
Therefore, $\tau^p_{s,0}$ are again equal for all $p$.
Dropping the processor index for $(\tau^p_{s,t})^2$, the recursion in (\ref{eq:MP_SE1}--\ref{eq:MP_SE3}) can be simplified as 
\begin{align*}
(\tau_{s,0})^2&=\sigma_W^2+\sum_{p=1}^P\kappa_p^{-1}\mathbb{E}\left[\left(\eta_{s,0}(X+\tau_{s,0}Z)-X\right)^2\right]\\
&=\sigma_W^2+\kappa^{-1}\mathbb{E}\left[\left(\eta_{s-1,0}(X+\tau_{s-1,0}Z)-X\right)^2\right],
\end{align*}
where the iteration evolves over $s$, which is identical to (\ref{eq:SP_SE}) evolving over $t$.

{\bf Numerical results for SE:} We provide numerical results for C-MP-AMP for the Gaussian matrix setting, where SE is justified rigorously. We simulate i.i.d. Bernoulli-Gaussian signals~\eqref{eq:BG} with $\rho=0.1$.
The measurement noise vector ${\bf w}$ has i.i.d. Gaussian $\mathcal{N}(0,\sigma_W^2)$ entries, where $\sigma_W^2$ depends on
the signal to noise ratio (SNR) as 
$\text{SNR}:=10\log_{10} \left((N\mathbb{E}[X^2])/(M\sigma_W^2)\right)$.
The estimation function $\eta_{s,t}$ is defined as $\eta_{s,t}(u)=\mathbb{E}[X|X+\tau^p_{s,t}Z=u]$, where $Z$ is a standard normal RV independent of $X$, and $\tau^p_{s,t}$ is estimated by $\|{\bf z}^p_{s,t}\|/\sqrt{M}$, which is implied by SE.
All numerical results are averaged over 50 trials.

Let us show that the MSE of C-MP-AMP is accurately predicted by SE when the matrix
${\bf A}$ has i.i.d. Gaussian entries with $A_{i,j}\sim\mathcal{N}(0,1/M)$.
It can be seen from Fig.~\ref{fig:SE} that the MSE 
achieved by C-MP-AMP from simulations (red crosses) matches
the MSE predicted by SE (black curves) at every outer iteration $s$ and inner iteration $t$ for various choices of numbers of inner iterations (the number of red crosses within a grid).

As discussed in Remark 1, the average estimation error of C-MP-AMP is no worse than that of AMP, which implies that C-MP-AMP can achieve the MMSE
of large random linear systems~\cite{GuoVerdu2005} when AMP achieves it.\footnote{AMP can achieve the MMSE in the limit of large linear systems when the model parameters 
($\kappa$, SNR, and sparsity of ${\bf x}$) are within a region~\cite{Krzakala2012probabilistic}.}
This is verified in Fig.~\ref{fig:MMSE}.

\section{Discussion}\label{sec:discussion}

This overview paper discussed multi-processor (MP) 
approximate message passing (AMP) for solving linear inverse
problems, where the focus was on
two variants for partitioning the measurement matrix.
In row-MP-AMP, each processor uses entire rows
of the measurement matrix, and decouples statistical 
information from those rows to scalar channels.
The multiple scalar channels, each corresponding to
one processor, are merged at a fusion center. We showed
how lossy compression can reduce communication requirements 
in this row-wise variant.
In column-MP-AMP, each node is responsible for 
some entries of the signal. While we have yet to consider
lossy compression in column-MP-AMP, it offers privacy
advantages, because entire rows need not be stored.
Ongoing work can consider lossy compression of 
inter-processor messages in column-MP-AMP,
as well as rigorous state evolution analyses.

\section*{Acknowledgments}
The authors were supported by the National Science Foundation (NSF) under grant ECCS-1611112. Subsets of this overview paper appeared in 
our earlier works, including in Han {\em et al.}~\cite{HanZhuNiuBaron2016ICASSP},
Zhu {\em et al.}~\cite{ZhuBeiramiBaron2016ISIT,ZhuBaronMPAMP2016ArXiv},
and Ma {\em et al.}~\cite{MaLuBaronICASSP2017}.
Finally, we thank Yanting Ma for numerous helpful discussions.

\bibliography{cites}

\end{document}